\documentclass[a4paper,fleqn,usenatbib]{mnras}

\usepackage{newtxtext,newtxmath}
\usepackage{graphicx}

\title[ULX masses from ultrafast outflows]{Constraining the mass of accreting black holes in ultraluminous X-ray sources with ultrafast outflows}

\author[Fiacconi et al.]{Davide Fiacconi$^{1,2}$\thanks{E-mail: fiacconi@ast.cam.ac.uk}, Ciro Pinto$^{1}$, Dominic J. Walton$^{1}$ and Andrew C. Fabian$^{1}$\\
$^{1}$Institute of Astronomy, University of Cambridge, Madingley Road, Cambridge CB3 0HA, UK\\
$^{2}$Kavli Institute for Cosmology, University of Cambridge, Madingley Road, Cambridge CB3 0HA, UK\\
}

\begin{document}

\date{\today}

\pagerange{\pageref{firstpage}--\pageref{lastpage}} \pubyear{2017}

\maketitle

\label{firstpage}


\begin{abstract}
The nature of ultraluminous X-ray sources (ULXs) -- off-nuclear extra-galactic sources with luminosity, assumed isotropic, $\gtrsim 10^{39}$~erg~s$^{-1}$ -- is still debated.
One possibility is that ULXs are stellar black holes accreting beyond the Eddington limit.
This view has been recently reinforced by the discovery of ultrafast outflows at $\sim$0.1-$0.2c$ in the high resolution spectra of a handful of ULXs, as predicted by models of supercritical accretion discs.
Under the assumption that ULXs are powered by super-Eddington accretion onto black holes, we use the properties of the observed outflows to self-consistently constrain their masses and accretion rates.
We find masses $\lesssim 100$~M$_{\sun}$ and typical accretion rates $\sim 10^{-5}$~M$_{\sun}$~yr$^{-1}$, i.e. $\approx 10$ times larger than the Eddington limit calculated with a radiative efficiency of 0.1.
However, the emitted luminosity is only $\approx 10\%$ beyond the Eddington luminosity, because most of the energy released in the inner part of the accretion disc is used to accelerate the wind, which implies radiative efficiency $\sim 0.01$.
Our results are consistent with a formation model where ULXs are black hole remnants of massive stars evolved in low-metallicity environments.
\end{abstract}



\section{Introduction}

Ultraluminous X-ray sources (ULXs) are non-nuclear, point-like, extragalactic sources with X-ray luminosity, assumed isotropic, $L_{\rm X} \gtrsim 10^{39}$~erg~s$^{-1}$; they are preferentially hosted by star-forming galaxies (see \citealt{feng+11} for a review).
There is no consensus yet on the nature of ULXs.
Indeed, their luminosities are larger than the Eddington luminosity of $\sim 10$~M$_{\sun}$ black holes (BHs) in Galactic binaries,
suggesting that they could represent a different class of objects, unless various combinations of significant beaming (e.g. \citealt{king+01}) and super-Eddington accretion (e.g. \citealt{begelman+02,poutanen+07,king+08}) are advocated.
On the other hand, ULXs could also be powered by sub-Eddington accretion onto intermediate-mass black holes (IMBHs) with masses $100 \lesssim M_{\bullet} / {\rm M}_{\sun} \lesssim 10^5$; however, IMBHs might be required only to explain the most luminous ULXs (e.g. \citealt{farrell+09}) and perhaps some of those showing quasi-periodic oscillations (e.g. \citealt{earnshaw+16}; but see also \citealt{middleton+11}), but they are unlikely to account for the majority of the ULX population (e.g. \citealt{stobbart+06,mapelli+08,gladstone+09}).
An intermediate possibility is represented by massive stellar BHs ($20 \lesssim M_{\bullet} / {\rm M}_{\sun} \lesssim 80$) produced by low metallicity ($Z \lesssim 0.4 Z_{\sun}$), massive ($\gtrsim 40$~M$_{\sun}$) stars \citep{mapelli+09,mapelli+10}, which are expected in ULX host galaxies owing to their low metallicity and high star formation rate \citep{pakull+02,zampieri+09}.
Moreover, some ULXs are powered by accretion onto neutron stars \citep{bachetti+14,furst+16,israel+16}; 
therefore, ULXs likely represent an heterogeneous class of objects.

Unfortunately, dynamical masses are not available for most ULXs.
Recently, \citet{liu+13} claimed the detection of the orbital modulation in M101 ULX-1, inferring a mass $>5$~M$_{\sun}$ and likely $\sim 30$~M$_{\sun}$.
However, the putative companion is likely a Wolf-Rayet star, which makes this dynamical measurement uncertain \citep{laycock+15}.
In addition to that, hard X-ray observations with NuSTAR of a few ULXs show a downturn in the spectra at energies $\gtrsim 10$~keV, which excludes sub-Eddington accretion onto IMBHs, whereas it favours super-Eddington accretion on lighter accretors \citep{bachetti+13,walton+14}.
If super-Eddington accretion is common among ULXs with luminosities $\sim 10^{40}$~erg~s$^{-1}$, an expected feature is the presence of optically-thick outflows from the inner, geometrically-thick, radiation-dominated regions of the accretion disc \citep{poutanen+07,takeuchi+13}.
Remarkably, recent work on high-resolution soft ($\sim$0.4-1.8~keV) X-ray spectra revealed for the first time blueshifted, high-excitation absorption lines compatible with velocity offsets as high as $\sim 0.25 c$ in 3 ULXs \citep{pinto+17,pinto+16}.
These results were strengthened by the high-energy counterpart of the NGC 1313 X-1 outflow observed in moderate-resolution broadband ($\sim$3-20~keV)
 X-ray spectra \citep{walton+16}.
The discovery of ultrafast outflows from ULXs strongly hints that at least a fraction of them may be powered by super-Eddington accretion.
However, as the nature of ULXs remains mostly unknown, it is unclear whether this would make them a peculiar class of accretors or an evolutionary phase of more common sources (e.g. high-mass X-ray binaries).

In this Letter, we assume that ULXs are BHs accreting beyond the Eddington limit and therefore that they are able to launch radiation-driven outflows from the inner part of the accretion disc.
We neglect here the case of ULXs powered by accreting neutron stars.
We use the observed properties of the outflows to self-consistently constrain the expected mass and accretion rate of the powering BH within such a framework.
Our results suggest that the few ULXs with observed outflows could be associated with BHs with masses between $\sim 10$ and $\sim 100$~M$_{\odot}$, and typical accretion rates $\sim 10^{-5}$~M$_{\sun}$~yr$^{-1}$.
If some ULXs are powered by BHs and ultrafast outflows caused by super-Eddington accretion are common, then our results suggest that their origin could be consistent with being the remnant of massive metal-poor stars.


\section{Supercritical accretion discs with outflows}

When the mass accretion rate within an accretion disc becomes supercritical, the excess of heat released by viscosity may both inflate the inner part of the disc, producing outflows, and be advected within the flow, increasing its entropy (e.g. \citealt{abramowicz+88}).
By supercritical we mean an accretion disc sustaining an accretion rate $\dot{M} > \dot{M}_{\rm Edd}$, where we define
\begin{equation}\label{eq_mdot_edd}
\dot{M}_{\rm Edd} \equiv  \frac{L_{\rm Edd}}{\tilde{\eta} c^2} = \frac{4 \pi G M_{\bullet}}{\tilde{\eta} \kappa_{\rm es} c} \approx 1.86 \times 10^{18}~m_{\bullet}~{\rm g~s^{-1}},
\end{equation}
where $M_{\bullet} = m_{\bullet}$~M$_{\sun}$ is the central BH mass, $\kappa_{\rm es}$ is the electron scattering opacity, and $\tilde{\eta} = 1/12$ is the (newtonian) radiative efficiency of a thin disc extending inward to 
$R_{\rm in} = 3 R_{\rm S} = 6 G M_{\bullet}/c^2$.
We stress that we adopt $\tilde{\eta}$ as a definition, and it can be different from the actual radiative efficiency $\eta \equiv L / (\dot{M} c^2)$ of a disc emitting $L$.

We base our analysis on the model of supercritical accretion disc with outflows presented by \citet{lipunova+99} and \citet{poutanen+07}.
The disc is composed of two regions separated by the spherisation radius $R_{\rm sp}$.
Outside $R_{\rm sp}$, the disc is geometrically thin and it can sustain a constant accretion rate $\dot{m} = \dot{M} / \dot{M}_{\rm Edd} > 1$ because it is locally sub-Eddington.
In this region, the local energy balance is $Q^{+} = Q_{\rm rad} + Q_{\rm adv} \approx Q_{\rm rad}$, where $Q^{+}$, $Q_{\rm adv}$, and $Q_{\rm rad}$ are the heat flux released by viscosity, the energy flux advected with the flow, and the energy flux that can eventually be radiated away, respectively.
The emitted luminosity in photons is $L_{\gamma}(R > R_{\rm sp}) = \int_{R > R_{\rm sp}} 2 Q_{\rm rad} {\rm d}A \approx L_{\rm Edd}$.
Inside $R_{\rm sp}$, the disc is locally super-critical, despite a significant energy fraction is advected with the flow.
In response, the disc becomes geometrically thick under the dominant effect of radiation pressure and a radiation-driven outflow mainly coupled with radiation through electron scattering is unavoidably launched.
Specifically, a fraction $\epsilon_{\rm w}$ of the available energy $Q_{\rm rad}$ is transferred to the kinetic luminosity of the wind, while the remaining can be radiated away.
The bolometric luminosity emitted by the disc is therefore
\begin{equation} \label{eq_en_cons}
L_{\gamma} \approx (1- \epsilon_{\rm w}) \int_{R_{\rm in}}^{R_{\rm sp}} 2 Q_{\rm rad} {\rm d}A + L_{\rm Edd} = \frac{1 - \epsilon_{\rm w}}{\epsilon_{\rm w}} P_{\rm w} + L_{\rm Edd},
\end{equation}
where we used the definition of the wind kinetic luminosity $P_{\rm w} = \epsilon_{\rm w} \int_{R_{\rm in}}^{R_{\rm sp}} 2 Q_{\rm rad} {\rm d}A$, and the additional $L_{\rm Edd}$ comes from the outer part of the disc.
The wind is radiation dominated and energy driven, and it is expected to be mainly accelerated at expense of the advected internal energy \citep{fiacconi+16}.
The conservation of energy expressed by equation (\ref{eq_en_cons}) already suggests an upper limit on the BH mass as $L_{\gamma} \geq L_{\rm Edd}$, which implies $M_{\bullet} \lesssim M_{\bullet, \rm Edd} \approx 72~l_{\gamma, 40}$~M$_{\sun}$, where $L_{\gamma} = l_{\gamma,40} \times 10^{40}$~erg~s$^{-1}$ and $M_{\bullet, \rm Edd}$ is the mass corresponding to $L_{\gamma} = L_{\rm Edd}$.
However, this limit is not general and applies only to sources where radiation-driven outflows are detected.

The spherisation radius $R_{\rm sp}$ determines the structure of the disc by setting where the transition between the outer geometrically thin and the inner geometrically thick disc occurs.
The value of $R_{\rm sp}$ depends on the mass, angular momentum, and energy conservation within the disc under the effect of viscosity, advection, and radiation, and it can be self-consistently calculated by solving the accretion disc equations.
\citet{poutanen+07} provide a fitting formula that depends on $\dot{m}$ and $\epsilon_{\rm w}$,
\begin{equation} \label{eq_r_mdot}
r_{\rm sp} \approx \dot{m} \left[1.34 - 0.4 \epsilon_{\rm w} + 0.1 \epsilon_{\rm w}^2 - \left(1.1 - 0.7 \epsilon_{\rm w} \right) \dot{m}^{-2/3} \right],
\end{equation}
where we have defined $r_{\rm sp} = R_{\rm sp} / R_{\rm in}$.

We can explicitly write the expression for $P_{\rm w}$ by considering the structure of the wind.
Within $R_{\rm sp}$ the scaling of the accretion rate is approximately linear,  ${\rm d} \dot{M} / {\rm d}R \approx \dot{M}_{\rm Edd} (\dot{m} - \dot{m}_{\rm in}) / r_{\rm sp}$, where $\dot{m}_{\rm in} = \dot{m}_{\rm in}(\epsilon_w, \dot{m})$ is the mass flow effectively reaching $R_{\rm in}$ in units of $\dot{M}_{\rm Edd}$ [see equation (23) of \citealt{poutanen+07}].
The rest of the mass accretion rate is accelerated in the outflow, which reaches an asymptotic velocity $v_{\rm w}(R) \approx \sqrt{2 G M_{\bullet} / R}$ within $R_{\rm sp}$.
Therefore, we finally obtain
\begin{equation} \label{eq_p_wind}
P_{\rm w} = \frac{1}{2} \int_{R_{\rm in}}^{R_{\rm sp}} \frac{{\rm d} \dot{M}}{{\rm d}R} v^2_{\rm w}(R) {\rm d}R \approx \frac{1}{2} \dot{M}_{\rm w} v_{\infty}^2 \frac{- \log(3 \beta_{\infty}^2)}{1 - 3 \beta_{\infty}^2},
\end{equation}
where we defined the mass outflow rate
\begin{equation}\label{eq_mdot_out}
\dot{M}_{\rm w} = \int_{R_{\rm in}}^{R_{\rm sp}} \frac{{\rm d} \dot{M}}{{\rm d}R} {\rm d}R \approx \dot{M}_{\rm Edd} (\dot{m} - \dot{m}_{\rm in}) \frac{r_{\rm sp} - 1}{r_{\rm sp}},
\end{equation}
and $\beta^{2}_{\infty} = v_{\infty}^2/c^2 = 1/(3 r_{\rm sp})$, i.e. $v_{\infty} = v_{\rm w}(R_{\rm sp})$.
The latter relation assumes that the radiative acceleration beyond $R_{\rm sp}$ is negligible; we checked that such simplification does not significantly affect our mass estimates below, while it tends to decrease the inferred accretion rate by a factor $< 1.5$.


\section{Constraining the black hole mass}

Recent observations have discovered fast outflows in the deepest X-ray spectra of ULXs (e.g. \citealt{pinto+16,walton+16}).
They are identified through high ionisation Fe, O, and Ne absorption lines produced by gas outflowing at $\approx 0.1$-$0.2 c$ with photoionisation parameters $\xi = L_{\rm ion} / (n r^2)$ ranging from $\sim 10^2$ to $\gtrsim 10^4$~erg~cm~s$^{-1}$, where $n$ is the density of the absorbing gas at radial distance $r$ from the source emitting the ionising luminosity $L_{\rm ion}$ in the energy band 1-1000~Ry.
We can use these results to constrain the parameters of a supercritical accretion disc potentially able to produce such winds and eventually the mass of the central BH as follows.

\begin{figure*}
\begin{center}
\includegraphics[width=2\columnwidth]{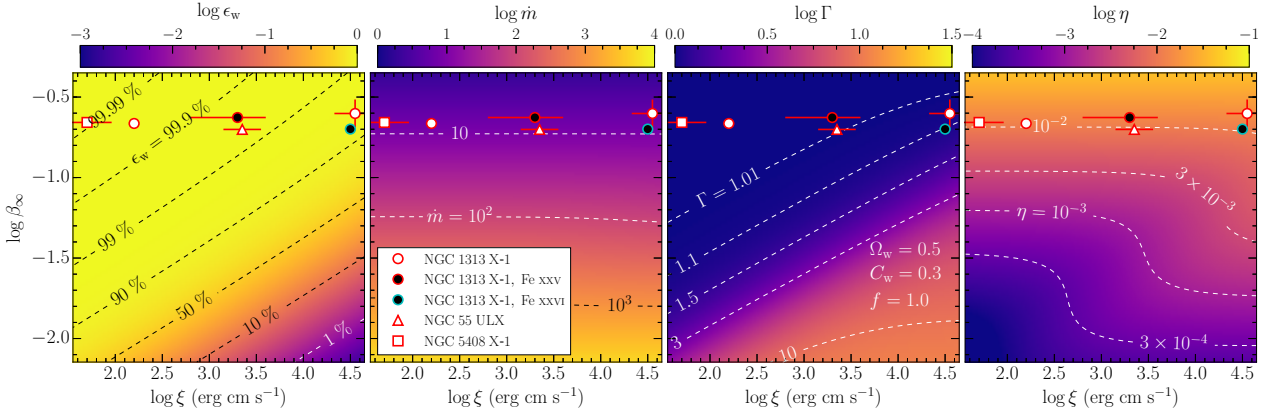}
\caption{
Results of the model shown in the $\xi$-$\beta_{\infty}$ plane.
From left to right: fraction $\epsilon_{\rm w}$ of the released energy within $R_{\rm sp}$ that is converted in kinetic luminosity of the wind,
accretion rate $\dot{m}$ in the outer part of the disc in units of $\dot{M}_{\rm Edd}$ [see equation (\ref{eq_mdot_edd})],
bolometric luminosity $\Gamma$ in units of $L_{\rm Edd}$,
and radiative efficiency $\eta = \tilde{\eta}~\Gamma/\dot{m}$.
We adopt the fiducial values for the model parameters: $\Omega_{\rm w} = 0.5$, $C_{\rm w} = 0.3$, and $f = 1$.
The symbols with error bars refer to the sources listed in Table \ref{tab_obs}: the white circles are the observations of NGC 1313 X-1 by \citet{pinto+16}; 
the black circles with red and cyan edges are the observations of NGC 1313 X-1 by \citet{walton+16} where the absorption feature is respectively attributed to \ion{Fe}{xxv} and \ion{Fe}{xxvi}; 
the triangle is the highest significance absorber in NGC 55 ULX detected by \citet{pinto+17}; and the square is the observation of NGC 5408 X-1 by \citet{pinto+16}.}
\label{fig_model_summary}
\end{center}
\end{figure*}

The conservation of mass in the outflow may simply read as
\begin{equation}\label{eq_mass_cons}
\dot{M}_{\rm w} \approx 4 \pi \Omega_{\rm w} r^2 C_{\rm w} \rho_{\rm w} v_{\infty},
\end{equation}
where we introduced a clumping factor $C_{\rm w}$ to phenomenologically capture the effect of a multi-phase wind and the fraction $\Omega_{\rm w}$ of the full sphere occupied by the outflow.
Under the simplifying assumption of a steady outflow, the density in the wind scales as $\rho_{\rm w} \propto r^{-2}$ at radii larger than $R_{\rm sp}$.
We can rewrite the density as a function of the photoionisation parameter $\xi$ as $\rho_{\rm w} = \mu m_{\rm p} L_{\rm ion} / (\xi r^2) \approx \mu m_{\rm p} f (L_{\gamma} - L_{\rm Edd}) / (\xi r^2)$, where $\mu \approx 0.59$ is the mean molecular weight of a fully ionised gas, $m_{\rm p}$ is the proton mass, and we express $L_{\rm ion} \approx f (L_{\gamma} - L_{\rm Edd})$,
which is the luminosity produced within $R_{\rm sp}$ that, after streaming through the wind beyond the wind photosphere, is able to photoionise the outflow, where $f \sim 1$ is an adjustable parameter to account for e.g. a small additional contribution of ionising photons from the outer disc or the conversion between bolometric and ionising luminosity.
If we introduce this definition in equation (\ref{eq_mass_cons}) and we then use it both in the equation of energy conservation, equation (\ref{eq_en_cons}), and in the definition of $\dot{M}_{\rm w}$, equation (\ref{eq_mdot_out}), we can solve for $\epsilon_{\rm w}$:
\begin{equation} \label{eq_solve_1}
\epsilon_{\rm w}(\xi, \beta_{\infty}) = \frac{2 \pi \Omega_{\rm w} C_{\rm w} f \tilde{\xi}^{-1} B(\beta_{\infty})}{1 + 2 \pi \Omega_{\rm w} C_{\rm w} f \tilde{\xi}^{-1} B(\beta_{\infty})},
\end{equation}
and we obtain
\begin{equation} \label{eq_solve_2}
\Gamma = 1 + \frac{\dot{m} - \dot{m}_{\rm in}(\epsilon_{\rm w}, \dot{m})}{4 \pi C_{\rm w} \Omega_{\rm w} f \tilde{\eta}}~\tilde{\xi}~\frac{1 - 3 \beta_{\infty}^2}{\beta_{\infty}},
\end{equation}
where $B(x) = -x^3 \log(3 x^2)/(1 - 3 x^2)$, and we normalise $\xi = \tilde{\xi} \mu m_{\rm p} c^3$ and $L_{\gamma} = \Gamma L_{\rm Edd}$.
The set of equation (\ref{eq_r_mdot}), (\ref{eq_solve_1}), and (\ref{eq_solve_2}), with the relation $r_{\rm sp} = 1/(3 \beta_{\infty}^2)$, fully characterises our problem.
After solving equation (\ref{eq_r_mdot}) numerically, we get $\epsilon_{\rm w}$, $\dot{m}$, and $\Gamma$ as a function of the observable quantities $\beta_{\infty}$ and $\xi$.
Then, we rescale $\dot{m}$ and $\Gamma$ to physical values through the definition of equation (\ref{eq_mdot_edd}) by choosing the value of the emitted bolometric luminosity $L_{\gamma}$, and we finally obtain $M_{\bullet}$ and $\dot{M}$ that are consistently required to have a supercritical accretion disc launching outflows at $v_{\infty}$.

\begin{figure}
\begin{center}
\includegraphics[width=\columnwidth]{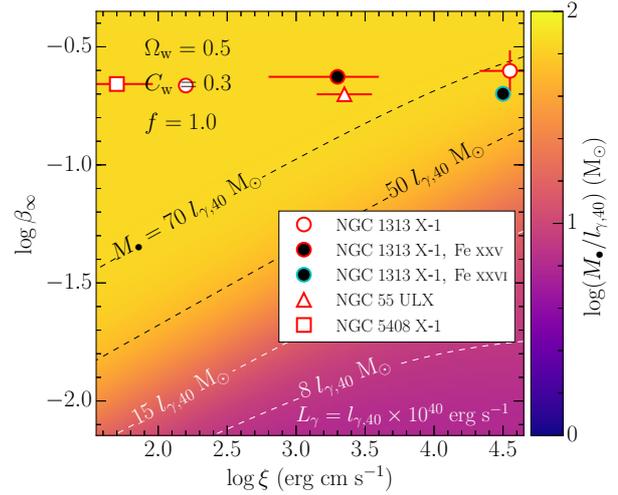}
\caption{
The colour-coded map shows the mass of the central BH scaled to a bolometric luminosity $L_{\gamma} = l_{\gamma,40} \times 10^{40}$~erg~s$^{-1}$.
We adopt the fiducial values for the model parameters: $\Omega_{\rm w} = 0.5$, $C_{\rm w} = 0.3$, and $f = 1$.
Observations are overplotted with the same symbols as in Figure \ref{fig_model_summary}.}
\label{fig_bh_mass}
\end{center}
\end{figure}

Figure \ref{fig_model_summary} summarises the results of our calculations, showing the values of $\epsilon_{\rm w}$, $\dot{m}$, $\Gamma$, and $\eta$ in the $\xi$-$\beta_{\infty}$ plane.
We adopt fiducial values for the parameters $\Omega_{\rm w} = 0.5$, $C_{\rm w} = 0.3$, and $f = 1$.
For values of $\beta_{\infty}$ above $\approx$0.05-0.1, the lines of constant $\epsilon_{\rm w}$ roughly scale as $\beta_{\infty} \propto [\epsilon_{\rm w} \xi / (1- \epsilon_{\rm w})] ^{1/3}$.
Mildly relativistic winds ($\beta_{\infty} \gtrsim 0.05$) typically require $\gtrsim 90\%$ of the energy produced within $R_{\rm sp}$ to accelerate the outflows, unless $\log \xi \gg 4$ and $\epsilon_{\rm w}$ can reduce to $\sim 50\%$.
The remaining energy is released in photons and it contributes to bolometric luminosities $\sim 10\%$ in excess of $L_{\rm Edd}$, as shown by the isocontours of $\Gamma$.
$\epsilon_{\rm w}$ can even exceed $99\%$ for fast outflows $\sim 0.2 c$ at $\log \xi < 3$, locking the bolometric luminosity at about $L_{\rm Edd}$.
On the other hand, the accretion rate $\dot{m}$ mainly depends on the outflow velocity because $\xi$ enters only weakly in equation (\ref{eq_r_mdot}) through $\epsilon_{\rm w}$ and $\dot{m} \sim r_{\rm sp} \sim \beta_{\infty}^{-2}$.
Fast winds with $\beta_{\infty} \gtrsim 0.1$ are typically associated with $\dot{m} \sim 10$, where $\sim 50 \%$ reaches the central BH while the remaining is accelerated in the outflows.
The differences between the isocontours of $\dot{m}$ and $\Gamma$ determines the actual radiative efficiency of the disc, $\eta = \tilde{\eta}~\Gamma / \dot{m}$, which is typically $\sim 0.001$ and increases up to $\sim 0.01$ for winds as fast as $\sim 0.2 c$.
This trend is in qualitative agreement with numerical simulations of supercritical accretion disc with radiation-driven outflows that show $\eta \sim 0.04$, larger than the equivalent slim disc \citep{jiang+14,sadowski+15}; however, we also infer $\eta$'s slightly lower than what numerical simulations predict.


\begin{table*}
\caption{Summary of the observational properties of ULXs with outflows and the model results.
From left to right: outflow velocity $\beta_{\infty}$, photoionization parameter $\xi$, X-ray luminosity $L_{\rm X}$ (assumed isotropic), estimated radiative efficiency $\eta$, estimated BH mass $M_{\bullet}$ (upper limits excluding beaming), estimated accretion rate $\dot{M}$.
The inferred $M_{\bullet}$ and $\dot{M}$ assume $L_{\rm X}$ as a proxy for $L_{\gamma}$.
}
\label{tab_obs}
\begin{tabular}{lccc ccc}
\hline
Source & $\beta_{\infty}$ & $\log \xi$ & $L_{\rm X}$ & $\eta$ & $M_{\bullet}$ & $\dot{M}$ \\
 & & (erg~cm~s$^{-1}$) & (erg~s$^{-1}$) &  & (M$_{\sun}$) & ($10^{-5}$~M$_{\sun}$~yr$^{-1}$) \\
\hline
NGC 1313 X-1$^a$ & $0.217 \pm 0.007$ & $2.20 \pm 0.04$ & $1.04 \times 10^{40}$ & $0.011$ & $75.0$ & $1.7$ \\
NGC 1313 X-1$^b$ & $0.25 \pm 0.05 $ & $4.55 \pm 0.22$ & '' & $0.015$ & $71.8$ & $1.2$ \\
NGC 1313 X-1$^c$ & $0.236 \pm 0.005 $ & $3.3^{+0.3}_{-0.5}$ & '' & $0.013$ & $74.8$ & $1.5$ \\
NGC 1313 X-1$^d$ & $\sim 0.2$ & $\sim 4.5$ & '' & $0.011$ & $68.4$ & $1.7$ \\
NGC 55 ULX$^e$ & $0.199 \pm 0.003 $ & $3.35 \pm 0.20$ & $1.3-2.1 \times 10^{39}$ & $0.009$ & $12.2$ & $0.3$ \\
NGC 5408 X-1$^f$ & $0.22 \pm 0.01 $ & $1.70 \pm 0.26$ & $2.01 \times 10^{40}$ & $0.012$ & $145.0$ & $3.1$ \\
\hline
\end{tabular}
\flushleft
Notes: $^{a}$ XABS 2 of Model 1 from \citet{pinto+16}; $^{b}$ XABS 3 of Model 2 from \citet{pinto+16}; $^{c}$ Absorption feature at $\approx 8.8$~keV attributed to \ion{Fe}{xxv} \citep{walton+16}; 
$^{d}$ Absorption feature at $\approx 8.8$~keV attributed to \ion{Fe}{xxvi} \citep{walton+16}; $^{e}$ Data for the most significant absorber at $\approx 3.5 \sigma$ from \citet{pinto+17}; $^{f}$ Data from \citet{pinto+16}.
\end{table*}

Figure \ref{fig_bh_mass} shows the corresponding BH mass in the $\xi$-$\beta_{\infty}$ plane for a source scaled to $L_{\gamma} = l_{\gamma, 40} \times 10^{40}$~erg~s$^{-1}$.
ULXs associated to outflows faster than $\sim 0.1 c$ are expected to be powered by BHs with masses $M_{\bullet} \sim 40$-$70~l_{\gamma, 40}$~M$_{\sun}$.
In fact, the isocontours follow those of the luminosity Eddington ratio $\Gamma$ and the mass asymptotically tends to the limiting mass $M_{\bullet, \rm Edd}$ when $\beta_{\infty}$ increases at constant $\xi$ (i.e. faster wind with higher $P_{\rm w}$), while it decreases when $\xi$ increases at constant $\beta_{\infty}$ (i.e. less dense wind with lower $P_{\rm w}$).
These trends arise because, for a given $L_{\gamma}$, a wind that has a larger $P_{\rm w}$ requires a higher fraction $\epsilon_{\rm w}$ of the energy produced within $R_{\rm sp}$, which implies that $\eta$ reduces and $\Gamma \rightarrow 1^{+}$, i.e. the emitted luminosity mostly comes from the outer, thinner disc.
We note that, for high $\beta_{\infty}$ and low $\xi$, our mass estimates tend to saturate at about $M_{\bullet, \rm Edd}$.

Both Figures \ref{fig_model_summary} and \ref{fig_bh_mass} show the positions in the $\xi$-$\beta_{\infty}$ of different observations of ULX outflows detected in three nearby star-forming low-metallicity spiral/dwarf galaxies.
The observations are summarised in Table \ref{tab_obs}; the table also shows the inferred radiative efficiency $\eta$, BH mass $M_{\bullet}$, and accretion rate $\dot{M}$.
All these quantities have been derived by assuming $L_{\gamma} = L_{\rm X}$, which could introduce a systematic error in the mass estimates.
However, if we take the results with caution as order of magnitude estimates, we see that our calculations constrain the BH masses for those ULXs grossly to range between $\sim 10$ and $\sim 100$~M$_{\sun}$, with accretion rates $\sim 10^{-5}$~M$_{\sun}$~yr$^{-1}$ (corresponding to $\dot{m} \sim 10$) and $\eta \sim 0.01$.

Finally, we note that the derived masses depend on three free parameters, namely $\Omega_{\rm w}$, $C_{\rm w}$, and $f$.
The fiducial values that we adopt for $C_{\rm w}$ and $\Omega_{\rm w}$, the latter corresponding to 60\degr from the rotation axes of the disc, are grossly expected from theoretical and numerical models \citep{king+08,takeuchi+13}.
The fudge factor $f$ is more uncertain, as it may account for different effects.
We tested the sensitivity of our results to these parameters by varying $C_{\rm w}$ and $\Omega_{\rm w}$ between 0.2 and 0.8, and $f$ between 0.2 and 3.
As they always appear together as $C_{\rm w} \Omega_{\rm w} f$ in equation (\ref{eq_solve_1}) and (\ref{eq_solve_2}), changing each of them independently has the same effect.
We find that $M_{\bullet}$ and $\dot{M}$ change by up to $\approx$15-20$\%$, i.e. they do not significantly affect our estimates.


\section{Discussion and conclusions} 

In this Letter, we attempted to constrain the mass and the accretion rate of 3 ULXs through the properties of their observed ultrafast outflows, under the assumption that the latter are caused by super-Eddington accretion onto BHs.
We find masses between $\sim 10$ and $\sim 100$~M$_{\sun}$ and accretion rates $\sim 10^{-5}$~M$_{\sun}$~yr$^{-1}$, about 10 times larger than $\dot{M}_{\rm Edd}$ in equation (\ref{eq_mdot_edd}).
However, the bolometric luminosity results to be only up to $\sim 10\%$ in excess of $L_{\rm Edd}$, implying a typical radiative efficiency $\eta \sim 0.01$, because $\gtrsim 90\%$ of the luminosity produced within $R_{\rm sp}$ is required to accelerate the outflow.

Bearing in mind that our small sample may not be representative of the whole population of ULXs, it is nonetheless interesting to note that the inferred masses lie between typical galactic binaries and the presumed low-mass tail of IMBHs (e.g. \citealt{casares+14}).
However, they should not be regarded as exotic; they are well consistent with the expected remnants of low-metallicity massive stars.
Indeed, \citet{mapelli+09,mapelli+10} have shown that such massive stellar BHs may statistically account for a large fraction of ULXs as well as for the correlation between the number of ULXs and the star formation rate of their host.
Therefore, this might support the speculation that many ULXs could be powered by super-Eddington accretion on $\sim 10$-100~M$_{\sun}$ BHs (e.g. \citealt{gladstone+09,middleton+15}).
However, a future larger sample of ultrafast outflows in ULX spectra as well as more robust dynamical mass estimates are necessary to eventually confirm this scenario.

Our investigation represents a tentative new approach to exploit recent observations of ultrafast outflows to constrain the mass of a few ULXs under the plausible assumption of super-Eddington accreting BHs.
However, we emphasise again that the derived masses should be taken as indicative.
Indeed, a factor of $\sim 2$-3 to reduce the derived masses may still be accommodated, because we have effectively neglected geometric beaming effects when we associate the bolometric luminosity to the observed one.
They might be particularly relevant for the radiation produced within $R_{\rm sp}$ that is mainly released in a rather narrow funnel along the disc rotation axis \citep{poutanen+07,takeuchi+13}.
As a consequence, our masses should be considered as upper limits because the true $L_{\gamma}$, and therefore $M_{\bullet}$, could be lower by the beaming factor $b \lesssim 0.5$-0.7 \citep{king+08,king+09}.
Moreover, we assumed $L_{\gamma} = L_{\rm X}$ for the sake of simplicity, but in fact the bolometric correction of the inferred X-ray luminosity might vary from system to system.
This might depend e.g. on the line of sight through the outflow, as hinted by the change in spectral hardness among the ULXs considered in Table \ref{tab_obs}, as well as by the conjectured connection with ultraluminous supersoft sources \citep{middleton+11,middleton+15,pinto+17}.
According to this scenario, NGC 1313 X-1 is likely to be seen more face-on because of the harder spectrum, and therefore $L_{\gamma} \approx L_{\rm X}$ is a reasonable assumption (but it is more likely to suffer from beaming though), while NGC 55 ULX and NGC 5408 X-1 have softer spectra that may come from the reprocessing of the harder radiation from the inner edge of the disc (and perhaps by a close hot corona) by the optically thick wind, for which $L_{\rm X} \lesssim L_{\gamma}$.
This latter correction might partially compensate the effect of beaming, but note that $L_{\rm X}$ as well is assumed isotropic and neglects beaming effects.
Nonetheless, even when reduced by a factor $\sim$2-3, our inferred masses are consistent with previous estimates of similar systems as well as with the aforementioned scenario of ULX formation.

Our calculations also neglect general relativistic effects, such as spinning BHs, and magnetic fields.
We explored the impact of a spinning BH by changing the inner boundary condition and the normalising radiative efficiency $\tilde{\eta}$.
We find values of $M_{\bullet}$ up to $\approx 20 \%$ lower than in the no spin cases, owing to larger values of $\Gamma$ and $\eta$ at similar accretion rates.
Magnetic fields might also contribute to accelerate the outflow \citep{blandford+82}, effectively lowering $\epsilon_{\rm w}$ and increasing $L_{\gamma}$ for the same accretion rate. 
This might also lower the value of the inferred central mass.

Regardless of their masses, BHs are not the only possible accretors powering ULXs.
Indeed, the light curves of 3 ULXs show sinusoidal pulses with a period $\approx 1$~s that are a distinctive signature of neutron stars \citep{bachetti+14,furst+16,israel+16}.
Recently, \citet{king+16} and \citet{king+17} argued that neutron stars may actually power a fraction of ULXs larger than previously expected.
According to their analysis, pulsations may only be observed during a rather short phase of the ULX evolution, preventing an easy detection.
However, the variability of the optical spectrum has revealed the existence of a massive stellar BH, likely $\sim 20$-30~M$_{\sun}$, in M101 ULX-1 \citep{liu+13} and perhaps also in X-ray binaries associated to Wolf-Rayet stars (\citealt{prestwich+07,crowther+10}; but see also \citealt{laycock+15}).
Therefore, it appears evident that ULXs comprise a diverse variety of accreting objects, and whether the majority is represented by BHs or neutron stars must be scrutinised further.
Clearly, our analysis can be applied only to the subset of ULXs with outflows that probably host an accreting BH; nonetheless, it would be possible to extend this treatment to neutron stars by modifying the inner boundary conditions in modelling the accretion disc owing to the magnetic fields.

Finally, we step on to more speculative grounds by noting that the BH masses we infer, when considered as upper limits as discussed above, are not too dissimilar from those of the binary BHs detected by Advanced LIGO \citep{abbott+16a,abbott+16b}.
While this might be just a coincidence, it is nonetheless worth to stress that such observations at least unambiguously demonstrate the existence in nature of rather heavy stellar BHs.
According to stellar evolution models, the most natural pathway to form $\gtrsim 30$~M$_{\sun}$ stellar BH is the evolution of massive stars in $Z < 0.1$-$0.5 Z_{\sun}$ environments \citep{belczynski+10,spera+15}.
Despite the evolution of binary stars is complicated by several physical processes that make difficult to predict the final outcome \citep{dominik+12}, it is still conceivable to imagine an evolutionary connection between binary massive stars, whose fraction can be as high as 70\% 
\citep{sana+08}, an intermediate phase when the binary turns into an ULX, and finally the merger of heavy stellar BH binaries (see also \citealt{pavlovskii+17}).
While this may sound attractive, further observations are required to better assess the puzzle of the nature of ULXs and consequently this potential connection.


\section*{Acknowledgements}

We acknowledge useful discussions with Massimo Dotti, Michela Mapelli, and Elena M. Rossi.
D.F. acknowledges support by ERC Starting Grant 638707 ``Black holes and their host galaxies: coevolution across cosmic time''.
C.P. and A.C.F. acknowledge support by ERC Advanced Grant 340442 ``Accreting black holes and cosmic feedback''.


\bibliographystyle{mnras}
\bibliography{ulx_outflows}


\label{lastpage}

\end{document}